# Multiple Energy Storage Rings

G. A. Krafft, B. Dhital, F. Lin, and V. Morozov
Jefferson Laboratory, Newport News, VA
Old Dominion University, Norfolk, VA

Abstract: Energy recovery linacs have been studied and developed over a number of decades. In the standard arrangement there is a separate particle source and particle beam dump. In this paper, a new arrangement is explored where the energy recovery linac is used as an energy source in a storage ring which has multiple beam energies. Several interesting topologies for the multiple beam energies are shown schematically. Possible energy equilibria with and without radiation damping are discussed. This idea may have applications to the problem of electron cooling.

In an energy recovery linac (ERL), an electron beam is recirculated through a linac so that the RF simultaneously accelerates and decelerates the beam [1,2]. Typically, this is accomplished by having the recirculating loop restricted to nearly an integer number of RF wavelengths, plus ½ a wavelength, and by sending the recirculated beam in the same direction as the accelerated beam through the linac. Within the linac, energy is transferred directly from accelerating beam to decelerating beam. The RF system must provide energy only for the current difference between the two beam loads.

Beam is accelerated to high energy in the ERL, used to some purpose, and then decelerated back to an energy that is close to the same as the initial beam energy where the beam is dumped. In the interest of obtaining maximum energy efficiency in an energy recovery linac, several authors have investigated whether it is possible to recover more energy before dumping the beam [3,4]. In this paper an alternative is suggested. Because the injection energy and dumped energy are close, there is no reason decelerated beam could not be fed back into an ERL at accelerating phase, again by having a beam line that is an integer plus ½ RF wavelength. When it is possible to close the beam orbit this way, the device looks like a storage ring with two separate beam energies. Such a two-energy storage ring has been discussed and developed in the context of advanced electron beam coolers [5] for the new Electron-Ion colliding beam accelerators [6,7]. Two-energy rings solve some of the difficult problems inherent in ERL-based coolers.

Recently, it has been realized that the two-energy ring idea can be further extended to construct multiple energy rings of at least four new topologies. The essential idea is that any number of intermediate energy recirculation loops can be added to the basic two-energy arrangement as long as the added recirculation loops have the proper total path length as measured in RF wavelengths, assuring the beam phase is unchanged by recirculation. Adding $m$ such loops yields an accelerator much like an $m+2$ energy storage ring. This idea is significant in that it may be a way to lower the cost of beam coolers by allowing shorter energy recovery linacs to achieve the same beam energy.

**Some topologies for potential multi-energy storage rings**

Conceptually, the solution easiest to build onto the standard two-energy ring is shown schematically in Figure 1. Loops at intermediate energies having an integral number of RF wavelengths total path length and accelerating energy recovery linacs are added as shown. To understand the simplest case, it is assumed that the transit time through each ½ loop above and below the linacs are the same. Starting from the lowest beam energy (loop 1) the attached first linac is phased to a certain accelerating voltage and synchronous phase, $\phi_s$. The beam traverses one half of loop 2 and arrives at linac 2. This linac could be phased to the same synchronous phase. Passing up through each succeeding ½ loop, phase each energy recovery linac to whatever voltage it produces, but at the same total synchronous phase. The highest energy loop introduces a 180 degree phase shift so on the second pass through linac $n-1$, perfect energy recovery occurs with the RF phase being at $\pi + \phi_s$. Because the transit time is equal on each side of the loop, as the particles encounter each linac on the way down in energy, the voltage phase is still $\pi + \phi_s$. This statement extends to the second pass through the first linac. The phasor describing the total RF voltage the particle sees after one circuit is given in Figure 2. After the total circuit the energy is back to where the particle started when synchrotron radiation is neglected.

Suppose now one of the linacs is phased to a different synchronous phase than the rest as in Figure 3. Because the highest energy loop introduces a common $\pi$ phase shift to all the linac voltages on second pass, including the one offset from the rest, it is still true that the resultant voltage of adding all accelerating and decelerating passes through the energy recovery linacs is zero. Thus one concludes that each of the energy recovery linacs can be phased to any voltage and any synchronous phase (even decelerating ones) and still have energy balance between the net accelerating and decelerating passes. Similarly, suppose the two halves of the loop at a given energy do not have the same transit time. Such a difference might lead to an offset of all the higher energy passes as seen in Figure 4, and a different resultant voltage. However, when the overall loop path length is still an integer, a shorter transit time through one half loop will be compensated by a longer transit time in the other half loop. The summed energy still balances, and can be corrected by an overall phase shift in the downstream linacs as above.

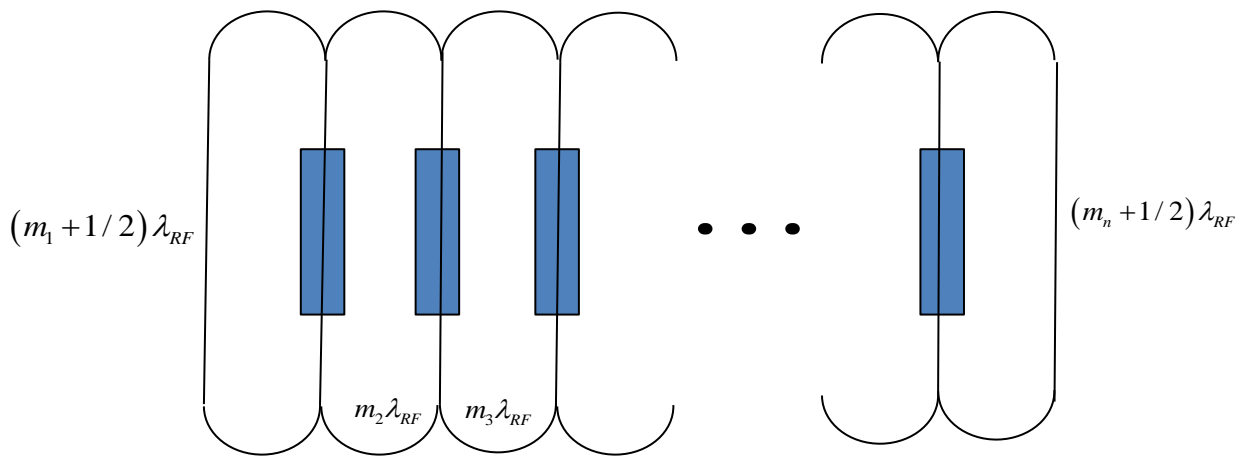

Figure 1: Multiple energy loop storage ring accelerator. Loops are labelled from 1 (at lowest energy) to loop n (highest energy) proceeding left to right

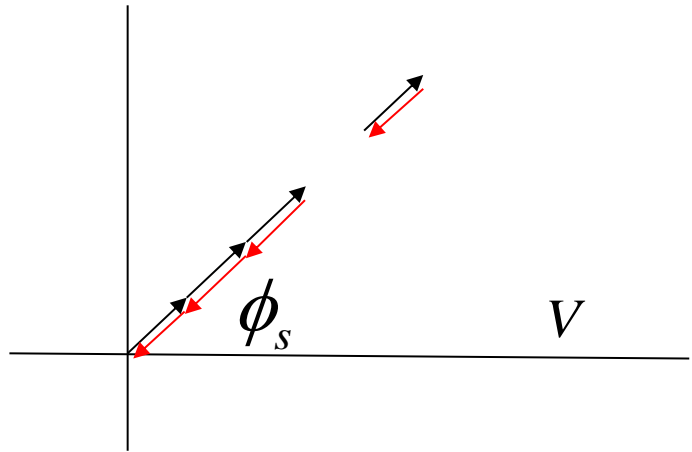

Figure 2: Accelerating phasor diagram for acceleration (black) and deceleration (red) in baseline arrangement. Red vectors displaced for clarity in discussion. They actually lie on top of, and cancel, the black vectors.

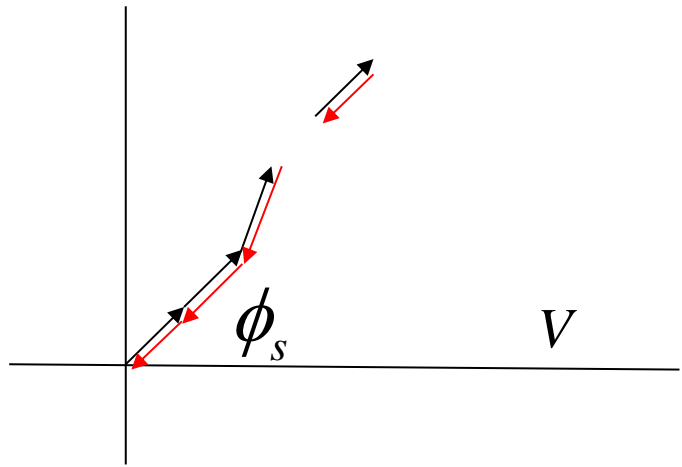

Figure 3: Same as Figure 2 with third linac phased differently. Still have overall cancelation in the sum.

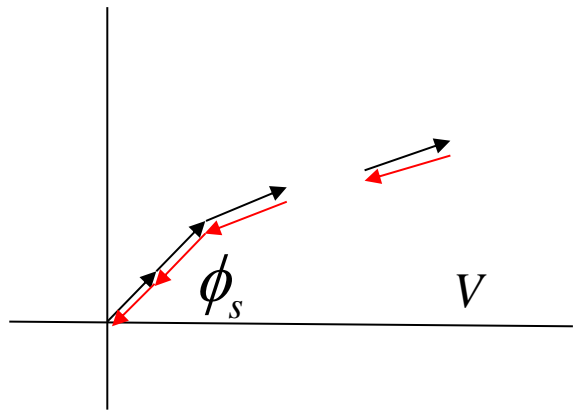

Figure 4: Same as Figure 2 with third linac phase offset because of a path length change in ½ of the third loop compensated by an opposite change in the other ½ loop. Still have overall cancelation in the sum.

This first topology provides a link between its lowest energy loop and its highest energy loop. However, the length of linac needed to go between these two energies is simply given by the energy difference divided by the linac average gradient, the same linac length needed in a simple two-energy ring. Referring to Figure 1, to reduce the length of linac needed to achieve a given energy difference, it would be highly beneficial to reuse the energy recovery linac structure as much as possible. A topology that accomplishes this goal is shown in Figure 5. As before, the lowest energy and the highest energy loops, which are transited only once in each acceleration/deceleration cycle, have an integer plus a half RF wavelengths. In the simplest rendering, intermediate energy beamlines need to be designed with a length of an integer number of RF wavelengths. All the intermediate energy beamlines will be transited twice each acceleration/deceleration cycle, once on the accelerating passes and once on the decelerating passes. There is much less flexibility in parameter choices. In particular, if one would like to vary the RF phases of the different beam passes by varying the length in the arcs, only those phase combinations having

$$\sum_{i=1}^{n} \Delta L_i = m\lambda_{RF},$$

where $m$ is an integer, are allowed.

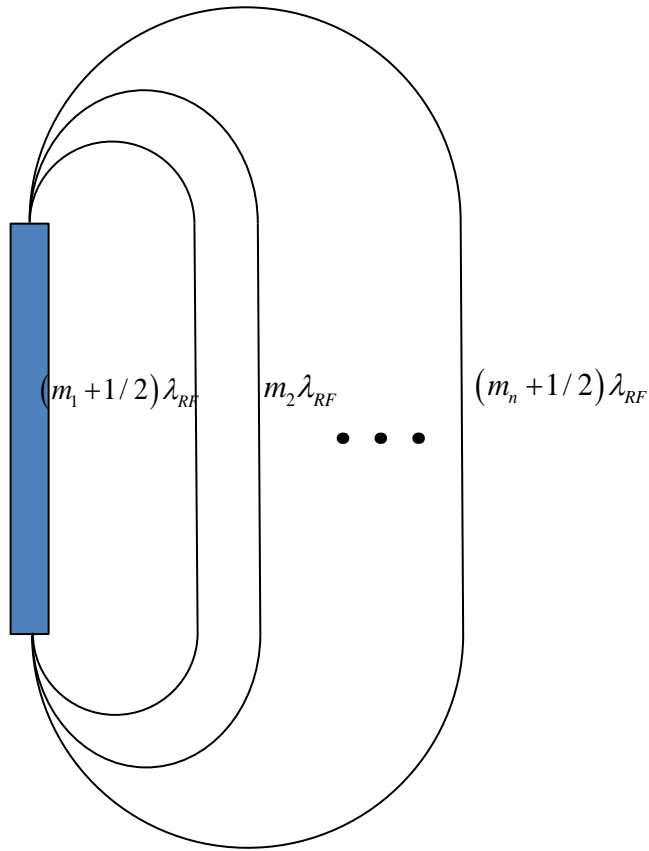

Figure 5: Multiple energy loop storage ring accelerator reusing the RF

A similar idea can be applied in "dog-bone" recirculation lattices where the accelerating and decelerating beam passes through the linacs in opposite directions [1,8,9]. As shown in Figure 6, in this case, to set up energy recovery, the highest energy arc and the lowest energy arc have an integer number of RF wavelengths, and the intermediate energy arcs now all have an integer plus ½ RF wavelengths. In the diagram a three-energy version is shown, where $l, m, n$ are integers. Any number of intermediate energy arcs are allowed.

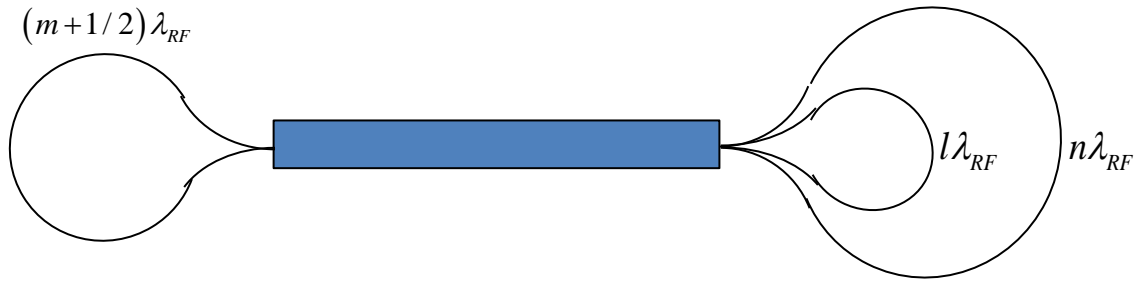

Figure 6: Multiple energy loop storage ring in "dog-bone" configuration

For cooling applications, it is very inconvenient that the lowest energy loop is on the inside of the accelerator in the second topology. Certainly, one can solve this problem by bending the low-energy loop out of the accelerator main plane to introduce the electrons into the ions. Maybe better, for those used to the CEBAF linacs, is the solution in Figure 7 for the 5-energy case. Here $l, m, n$ are integers. Add as many integer passes as desired. One half integer differences must appear on the highest energy arc and the lowest energy arc. By extension in the obvious way, this idea can be used to make a storage ring of any polytron accelerator arrangement [10].

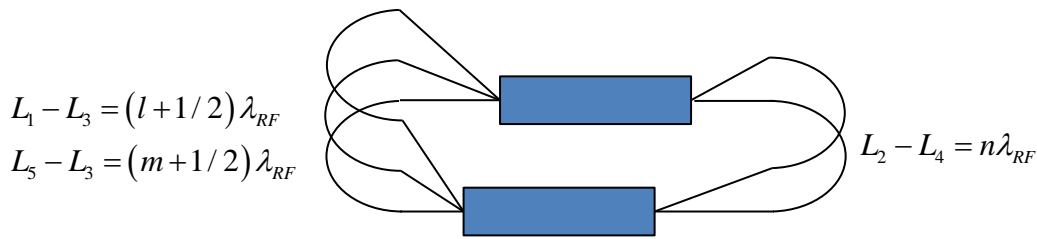

Figure 7: CEBAF-like multiple energy storage ring

These ideas may be tested at existing accelerators. For example, the second topology has great resemblance to the MAMI racetrack microtrons [11]. One of the microtrons could be made into a multiple energy storage ring by including ½ wavelength chicanes in their highest and lowest energy recirculation loops, and by providing for beam injection using the lower energy racetrack microtron. Similarly, CBETA [12] could be made into a multiple energy ring. Jefferson Lab has plans to perform a multipass energy recovery experiment by adding a ½ wavelength chicane in the highest energy arc [13]. Once this modification is performed, a multiple energy storage ring is easily made by adding a ½ wavelength chicane in its lowest energy ARC, and by adding needed beam injection components. Not surprisingly, close variants of these basic topologies have been studied in reference to ERL-driven light sources [14-18].

**Equilibria with Synchrotron Radiation**

Up to this point the effect of synchrotron radiation has not been corrected. The energies after each stage of acceleration and deceleration are equal in the geometries considered. When the effects of synchrotron radiation are included, this situation is no longer true. Two methods may be used to couple net energy into the beam to compensate for synchrotron radiation. The conceptually simplest method is to add a non-recovered RF cavity into either the lowest energy loop or highest energy loop, either of which support only single beam passes. An (over)voltage is placed on the cavity beyond the energy loss per turn. As in a single energy storage ring the beam phase will adjust so that

$$\Delta E = V_{ov} \cos \phi_s$$

where $\Delta E$ is the energy loss per turn from the radiation, $V_{ov}$ is the overvoltage, and $\phi_s$ is the synchronous phase of the beam with respect to the extra cavity referenced to the voltage crest phase. As long as the extra cavity is phase locked with all the energy recovery linacs in the topology, the phase of the beam with respect to any of the ERLs can be adjusted by introducing offset phases into the RF controls of the ERLs. Of course, arbitrary offset phases are not necessarily allowed. Only those sets consistent with the energy gain program of the loop designs can be used.

At the risk of somewhat greater complications, the extra cavity is not necessary. Imagine the beam acceleration program in Figure 8. The accelerating phase is given by $\phi_s$ and by introducing a path length change in the highest energy loop and a negative compensating path length change in the lowest energy arc, the decelerating phase will be offset with respect to the accelerating phase. As long as the vector sum of acceleration and deceleration has an amplitude beyond the synchrotron energy loss, the beam will settle down to a phase such that

$$\Delta E = V_{net} \cos \phi_{net}$$

where $V_{net}$ is the amplitude of the summed voltage and $\phi_{net}$ is the beam phase with respect to this net voltage. Usually the amplitude of $V_{net}$ is much less than the amplitude of the total accelerating (or decelerating) voltage. In this case, by performing the simple trigonometry, one notes that

$$\phi_s - \phi_{net} \approx \pi/2.$$

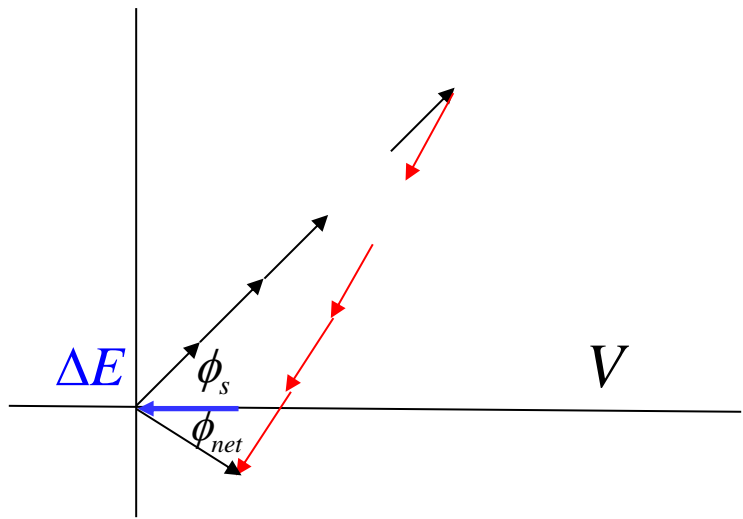

Figure 8: Phasor addition when highest energy loop path length is adjusted, with a compensating adjustment of the lowest energy loop. The phase of acceleration and deceleration are offset. Offsets not to scale and enhanced for clarity of presentation. Energy loss from synchrotron radiation shown in blue.

An addition complication is that to achieve a given $\phi_s$ in the linacs, the amplitude of the net voltage must be "tuned" by proper choice of the path length change so that correct energy gain is achieved at the desired offset phase. Also, now the energies of the beam in the intermediate loops are slightly offset from each other.

One might worry that the net beam phase, being negative, is wrong for synchrotron oscillation stability. In another paper we have analyzed this problem and shown that synchrotron stability is in fact dominated by the effects of the energy recovery linacs [19]. As long as the linac phases are synchrotron stable, there will be significant synchrotron stability and a very large longitudinal acceptance in the overall accelerator system [20].

**Summary**

In this paper it is shown that the idea of a two-energy storage ring can be extended to a multiple-energy storage ring by adding intermediate energy loops that have fixed recirculation path length. Several different possible topologies yielding multiple beam energies have been presented. Stability against synchrotron emission can be assured by either addition of an RF cavity in the non-recovered beam lines in the accelerator or by shifting the decelerating phase away from the accelerating phase by appropriate path length adjustments in the highest energy loop and lowest energy loop.

**Acknowledgement**

The authors thank Todd Satogata for a pertinent question that stimulated them to investigate the possibilities of multiple energy rings. This manuscript has been authored by Jefferson Science Associates, LLC under Contract No. DE-AC05-06OR23177 with the U. S. Department of Energy. The United States Government retains and the publisher, by accepting the article for publication, acknowledges that the United States Government retains a non-exclusive, paid-up, irrevocable, world-wide license to publish or reproduce the published form of this manuscript, or allow others to do so, for United States Government purposes. Additional support for this work was provided by the Office of Nuclear Physics, United States Department of Energy, under Contract No. DE-AC02-06CH11357.

# References

[1] M. Tigner, "A possible apparatus for electron clashing-beam experiments", Il Nuovo Cimento, **37**, 1228 (1965)

[2] L. Merminga, D. R. Douglas, and G. A. Krafft, "High-Current Energy-Recovering Electron Linacs", Annual Reviews of Nuclear and Particle Science, **Vol. 53**, 387-429 (2003)

[3] F. Marhauser, R. Rimmer, F. Hannon, and R. R. Whitney, "Method for Energy Recovery of Spent ERL Beams", United States Patent 9,872,375 issued Jan. 16, 2018

[4] R. Ainsworth, G. Burt, I. V. Konoplev, and A. Seryi, "Asymmetric dual axis energy recovery linac for ultrahigh flux sources of coherent x-ray and THz radiation: investigations towards its ultimate performance", Phys. Rev. AB **19**, 083502 (2016)

[5] F. Lin, Y. S. Derbenev, D. Douglas, J. Guo, G. A. Krafft, V. S. Morozov, Y. Zhang, and R. P. Johnson, "Storage-ring Electron Cooler for Relativistic Ion Beams", Proc. of the 2016 Int. Part. Acc. Conf., Busan, Korea, 2466 (2016)

[6] Y. Zhang, "JLEIC: A High Luminosity Polarized Electron-ion Collider at Jefferson Lab", Proc. of the 2019 Int. Part. Acc. Conf., Melbourne, Australia, 1916 (2019)

[7] C. Montag, "eRHIC Design Overview", Proc. of the 2019 Int. Part. Acc. Conf., Melbourne, Australia, 45 (2019)

[8] S. A. Bogacz, "Multi-pass ERL in a 'Dogbone' Topology", Proc. of the 2019 Int. Part. Acc. Conf., Melbourne, Australia, 1601 (2019)

[9] S. A. Bogacz, 'Muon Acceleration Concepts for NuMAX:"Dual-use" Linac and "Dogbone" RLA', Jour. of Inst., **13**, P02002 (2018)

[10] R. E. Rand, *Recirculating Electron Accelerators*, Harwood Academic Publishers, New York, (1984)

[11] A. Jankowiak, "The Mainz Microtron MAMI – Past and Future", Eur. Phys. Jour. A – Hadrons and Nuclei **28** (Suppl 1), 149 (2006)

[12] G. H. Hoffstaetter, et al., "CBETA Design Report, Cornell-BNL ERL Test Accelerator", arXiv:1706.04245 [physics.acc.ph], June 2017

[13] F. Méot, et al., "ER@CEBAF - A High Energy, Multi-pass Energy Recovery Experiment at CEBAF", Proc. of the 2016 Int. Part. Acc. Conf., Busan, Korea, 1022 (2016)

[14] C. Wang, J. Noonan, and J. W. Lewellen, "Dual-axis Energy-recovery Linac", Proc. of the 2007 ERL Conf., Daresbury, UK, 122 (2007)

[15] A. N. Matveenko, T. Atkinson, A. V. Bondarenko, and Y. Petenev, "Feasibility Study of Multi-turn ERL-based Synchrotron Light Facility", Proc. of the 2013 ERL Conf., Novosibirsk, Russia, 80 (2013)

[16] T. Atkinson, A. N. Matveenko, A. V. Bondarenko, and Y. Petenev, "Conceptual Design Report for a multi-turn Energy Recovery Linac-based Synchrotron Light Facility (Femto-Science Factory)", Helmholtz-Zentrum Berlin, August (2014)

[17] G. H. Hoffsteatter, S. M. Gruner, and M. Tigner, editors, "Cornell Energy Recovery Linac: Project Definition Design Report", Cornell Univerisity, June (2013)

[18] S. Benson, D. Douglas, Y. Roblin, T. Satogata, T. Powers, C. Tennant, R. York, D. Angal-Kalinin, N. Thomson, A. Wheelhouse, P. Williams, and T. Charles, "Architectural Considerations for Recirculated and Energy Recovered Hard XFEL Drivers", Proc. of the 2018 Int. Part. Acc. Conf., Vancouver, BC, Canada, 4560 (2018)

[19] B. Dhital, G. A. Krafft, and F. Lin, "Comments on Equilibria and Synchrotron Stability in Storage Ring Coolers", unpublished

[20] B. Dhital, J. R. Delayen, Y. S. Derbenev, D. Douglas, G. A. Krafft, F. Lin, V. S. Morozov, and Y. Zhang, "Equilibria and Synchrotron Stability in Two Energy Storage Rings", Proc. of the 2019 Int. Part. Acc. Conf., Melbourne, Australia, 364 (2019)